# Cosmic Ray Acceleration by Stellar Associations? The Case of Cygnus OB2

Yousaf Butt[1], Paula Benaglia[2], Jorge Combi[2], Michael Corcoran[3], Thomas Dame[1], Jeremy Drake[1], Marina Kaufman Bernadó[2], Peter Milne[4], Francesco Miniati[5], Martin Pohl[6], Olaf Reimer[6], Gustavo Romero[2], Michael Rupen[7]

[1]*Harvard-Smithsonian Center for Astrophysics, 60 Garden St., Cambridge, MA, USA*

[2]*Instituto Argentino de Radioastronomía, CC 5, 1894, Villa Elisa, Buenos Aires, ARGENTINA*

[3] *Universities Space Research Association, 7501 Forbes Boulevard, Suite 206, Seabrook, MD 20706 and Laboratory for High Energy Astrophysics, Goddard Space Flight Center, Greenbelt, MD 20771*

[4]*Los Alamos National Laboratory (T-6),Mail Stop B-277,  Los Alamos, NM 87545, USA*

[5]*Max-Planck-Institut für Astrophysik, Karl-Schwarzschild-Strasse 1, D-85741 Garching, GERMANY*

[6] *Institut für Theoretische Physik, Ruhr-Universität Bochum, 44780 Bochum, GERMANY*

[7] *National Radio Astronomy Observatory, Socorro, NM, USA*

**The origins of all extra-solar cosmic rays – both the ultra high-energy and lower energy Galactic variety – remain unclear. It has been argued that the large scale shocks and turbulence induced by the multiple interacting supersonic winds from the many hot stars in young stellar associations may play a role in accelerating Galactic cosmic rays –** *with or without the associated multiple supernova remnants* **(eg. Cesarsky & Montmerle, 1983; Bykov 2001). In this context, the report by the HEGRA Cherenkov telescope group of a steady and extended unidentified TeV gamma-ray source lying at the outskirts of Cygnus OB2 is particularly significant. This is the most massive stellar association known in the Galaxy, estimated to contain ~2600 OB type members alone – indeed, Cyg OB2 also coincides with the non-variable MeV-GeV range unidentified EGRET source, 3EG 2033+4118. We summarise[1] here the near-simultaneous follow-up observations of the extended TeV source region with the CHANDRA X-ray Observatory and the Very Large Array (VLA) radio telescope. Applying a numerical simulation which accurately tracks the radio to gamma-ray emission from primary hadrons as well as primary and secondary $e^{\pm}$, we find that the broadband spectrum of the TeV source region favors a predominantly nucleonic – rather than electronic – origin of the high-energy flux, though deeper X-ray and radio observations will help confirm this. A very reasonable, ~0.1%, conversion efficiency of Cyg OB2's extreme stellar wind mechanical luminosity to nucleonic acceleration to ~PeV ($10^{15}$ eV) energies is sufficient to explain the multifrequency emissions.**

---

[1]*This is a condensed, conference proceedings version of a longer manuscript, astro-ph/0302342.*

I. INTRODUCTION

There is currently a great deal of interest in identifying the possible sources of the ultra-high energy cosmic rays. However, it should be emphasized that the astrophysical sites where even the lower-energy (up to ~$10^{15}$ eV/nucleon) Galactic cosmic ray (GCR) nuclei are accelerated continue to defy identification. The expanding shock waves of supernova remnants (SNRs) have long been conjectured to be the accelerators of GCRs based mostly on energetic and spectral consistency arguments (eg. Ginzburg & Syrovatskii 1969; Drury et al., 2001). Recent observations from ground-based Cherenkov gamma-ray telescopes have provided direct evidence of TeV range *electrons* in individual SNRs (eg. Muraishi et al, 2000), although the situation for nuclei remains more confused (eg. Reimer & Pohl, 2002; Butt et al 2002; Torres et al., 2003; Erlykin & Wolfendale 2003).

Since most SNe explosions are core-collapse SNe of massive progenitors (M≳$8M_\odot$), and since such progenitor stars are typically formed in associations, it is plausible that the resultant 'superbubbles' (Heiles, 1979) – characterized by the collective shocks induced by close-by and time-correlated SN explosions – should be even more promising GCR source sites. For recent reviews see, eg., Bykov (2001) and Parizot (2002). Cesarsky & Montmerle (1983) demonstrated how the turbulent interacting supersonic stellar winds of the many young OB stars in some associations could dominate the GCR acceleration process for the first 4-6 Myrs, even before the first SNe begin to explode. In fact, they suggested that such 'cumulative' OB association stellar winds may be even more efficient than individual SNRs in accelerating GCRs for two reasons: the stellar wind shocks will be turbulent on both sides of the shock interface (thus speeding up the acceleration process); and, since there is continuous energy input, the shock velocity can remain higher for longer than in the impulsively powered SNR shocks.

In this context, the recent report by the HEGRA collaboration of an extended and steady TeV source within the boundary of the Cyg OB2 stellar association (Rowell et al., 2002; Aharonian et al. 2002; Horns & Rowell, 2003) provides an ideal opportunity to test the stellar association hypothesis of GCR origin. The low latitude of the source, its ~11 arcmin (Gaussian best-fit) extension, and lack of variability, all point to a Galactic origin[2]. At (4-10) ×$10^4$ $M_\odot$, Cyg OB2 is the most massive OB association known in the Galaxy; the reader is referred to, eg., Reddish, Lawrence & Pratt (1966); Knödlseder (2000); Comeron et al. (2002); Uyaniker et al. (2001); and, Knödlseder (2002) for useful overviews. Though it houses some of the most massive and luminous stars in the Galaxy – including the only two extreme O3 If* type stars known in the northern hemisphere (stars 7 and 22-A; Knödlseder, 2002) – Cyg OB2 is also a rather compact association: at 1.7 kpc it has a diameter of ~60 pc, or ~2°. This implies a tremendous mechanical power density from the cumulative stellar winds of its ~2600 OB star members: Lozinskaya et al. (2002) estimate that an average of a few $10^{39}$ erg/sec must have been continuously released over the past ~2Myrs in this region. That the non-variable gamma-ray source, 3EG J2033+4118 (2EG J2033+4112/GRO J2032+40) (Hartman et al. 1999), was found to be centered on Cyg OB2 argues strongly in favor of a physical association (White & Chen, 1992; Chen & White, 1996).

In Figure 1 we show the stellar density plots of all *cataloged* OB member stars together with overlays indicating the positions of 3EG J2033+4118 and TeV J2032+4131 – interestingly, the TeV source coincides with a distinct sub-group of outlying OB stars. Note that many stars in Cyg OB2 remain undetected and uncataloged due to high visual extinction in this direction (eg. Comeron et al., 2002). Six cataloged O, and eight cataloged B stars lie within the reported extent of the TeV source, but again these numbers should be considered strict lower limits. Their parameters are detailed in Table 1.

II. Observations

a. CHANDRA

We obtained a 5 ksec Director's Discretionary Time (DDT) CHANDRA observation of TeV J2032+4131 ($\alpha_{2000}$: $20^{hr}32^{m}07^{s}\pm9.2^{s}\pm2.2^{s}$, $\delta_{2000}$: +41°30′30″±2.0′±0.4′, radius~5.6′; Aharonian et al., 2002) starting on 11 August 2002 19:51 GMT (OBSID 4358). The details of the analysis may be found in Butt et al. (2003). Figures 2 & 3 show the raw and smoothed X-ray images.

---

[2] However, the extragalactic alternative cannot be altogether eliminated: an extended *extragalactic* TeV source, the starburst galaxy NGC 253, has been recently reported by the CANGAROO collaboration (Itoh et al., 2002; Itoh et al., 2003) and a possible explanation in terms of cosmic rays illuminating the core regions of massive stars there has been put forth by Romero & Torres (2003) [see also, Anchordoqui, Romero, and Combi, 1999].

Unfortunately, we found that due to the low statistics obtained, the residual TeV source region X-ray spectrum could be equally well-represented by optically-thin plasma models (the *MEKAL* model) or non-thermal power laws. Similar reduced $\chi^2$ values of ~0.9 were obtained for both models. Based on the best-fit spectral models, we obtain a diffuse flux within the source region of $1.3 \times 10^{-12}$ ergs cm$^{-2}$ sec$^{-1}$ for the 0.5-2.5 keV bandpass, and $3.6 \times 10^{-12}$ ergs cm$^{-2}$ sec$^{-1}$ for the 2.5-10 keV bandpass. *Unfortunately, because both power law and thermal plasma models are equally acceptable, the flux values extracted above may only be taken as upper limits to the non-thermal component alone.* Consequently, in our quantitative modeling (Section III) of the multiwavelength emissions we have taken the measured (instrumental background subtracted) X-ray flux as an upper limit to the X-ray emission associated with the TeV source. A deeper, ~50 ksec, observation would yield sufficient counts to permit a reliable decomposition of the X-ray emission into thermal and power-law components.

Spectra were also extracted for different regions surrounding the TeV source region, including the brighter region to the southeast. The TeV source region showed no significant excess hardness compared to these other regions and spectra were qualitatively very similar.

b. VLA B- and D-configuration Observations

On the following day, 12 August 2002 we obtained a 8 minute 4.86 GHz VLA[3] exposure in the B-configuration, sampling a 10.24′ ×10.24′ region centered at the TeV source (the half-power sensitivity region of the antenna is about 9′ diameter in this configuration). In the B-configuration, the VLA array is sensitive only to point-like radio sources. We achieved an rms noise of 96$\mu$Jy/beam for a beam size (psf) of 1.50″ ×1.42″ (FWHM), oriented 28° E of N. We detected no point-like sources to the limiting flux in the region of interest sampled by the primary beam.

Since the VLA B-configuration data we obtained is not sensitive to any possible diffuse radio emission present in the TeV source region, we reanalyzed archival D-configuration data at 1.489 GHz taken in 1984 from which we obtained an upper limit to diffuse emission of <200mJy in the region of the TeV source (Figure 4). Our analysis (Section III) assumes no time variability of the source since 1984, consistent with the multi-year steadiness reported by HEGRA.

e. EGRET

The >100 MeV source, 3EG J2033+4118, whose 95% and 99% confidence location contours overlap the extended TeV source region (Fig 1), is a ~12$\sigma$ detection centered at $l$=80.27º, $b$=+0.73º, with a radial positional uncertainty $\theta_{95\%}$=0.28º (Hartman et al. 1999). An elliptical fit by Mattox, Hartman & Reimer (2001) yields the parameters $a$=18.7', $b$=15.0', $\phi$=67º, where $a$ and $b$ are the length of the semimajor and semiminor axes in arcmin, and $\phi$ is the position angle of the semimajor axis in. 3EG J2033+4118 is classified as being a non-variable source by Tompkins (1999), Torres et al. (2001), McLaughlin et al. (1996; $V$=0.61 for 2EG J2033+4112) and M. McLaughlin ($V$=0.4 for 3EG J2033+4118; personal comm., 2003). Bertsch et al. (2000) and Reimer & Bertsch (2001) concluded, that in the case of 3EG J2033+4118 a double power law spectral fit or a power law fit with exponential cutoff are more appropriate than a single power-law. This could partially explain the discrepancy between the EGRET flux and the HEGRA flux in a spectral energy distribution (see Fig 3 in Aharonian et al. 2002) – *if the MeV/GeV emission and the newly discovered TeV source are indeed directly related to the same astronomical object in the Cygnus region*. However, such a scenario is highly problematic in that after the index softens in the GeV range it would then have to re-harden to ~ −1.9 at the TeV energies observed by HEGRA. In our opinion, such an interpretation appears to be overly contrived.

Thus, while 3EG J2033+4118 and GeV J2035+4214/GRO J2034+4203 may be due to the same object(s), it is unlikely that the TeV source is *directly* related to any of them. 3EG J2033+4118 is probably connected with some subset of the ~2600 OB stars in the core of Cyg OB2, whereas TeV J2032+4131 could be related to the region coincident with an outlying OB sub-group as shown in Fig 1. The sources may, however, still be considered indirectly related if the particles accelerated to GeV energies by the cumulative wind-shocks from the Cyg OB2 core stars, are reaccelerated to TeV energies by the collective wind shocks and turbulence in the region of the outlying OB sub-group. Verifying such a scenario will require deeper multiwavelength observations.

---

[3] The VLA is operated by the National Radio Astronomy Observatory (NRAO), which is a facility of the National Science Foundation (NSF), operated under cooperative agreement by Associated Universities, Inc. (AUI).

b. 60 & 100$\mu$m IRAS emission

An examination of the reduced 60 & 100$\mu$m IRAS data (eg. Fig 4b in Odenwald & Schwartz 1993 and Fig 1 in Le Duigou & Knodlseder 2002) clearly shows a dust void at the location of the TeV source. Odenwald & Schwartz (1993) argue that this void is due to the violent stellar environment of Cyg OB2: either the dust has been evacuated from Cyg OB2 – and the TeV source region especially – or else it has been destroyed.

In summary, the molecular and dust maps show a low density region at the location of the TeV source, most plausibly due to the action of the massive core stars of Cyg OB2, as well as the outlying OB sub-group coincident with the TeV source (Fig 1). The co-added atomic+molecular+ionized density of the region of the TeV source is ~30 nucleons cm$^{-3}$ (Butt et al., 2003).

III. Modeling the Multifrequency emission.

Determining whether the TeV photons are dominantly produced by electronic or nuclear interactions is, of course, of fundamental importance in assessing whether Cyg OB2 may be considered a nucleonic GCR accelerator. In order to do this, we considered two main cases: one in which the TeV source is due predominantly to $\pi^\circ \rightarrow \gamma\gamma$ emission from interactions of energetic nucleons; and the other in which IC upscattering of CMB photons by relativistic electrons generates the bulk of observed gamma-rays. (Considering the measured density of the TeV source region, the IC process will outshine electronic bremsstrahlung in the TeV gamma-ray domain, so we are justified in considering just the two cases mentioned).

To do so we assume that the putative acceleration mechanism (either shock and/or turbulent acceleration) generates a power-law spectrum of primary particles with a normalization, slope and maximum energy chosen to agree with those determined empirically from the observed TeV spectrum. Following Aharonian et al. (2002) we take the spectral index as −1.9 and the maximum particle energy as 1 PeV. The required kinetic energy of the injected particles corresponds to only a fraction of a percent of the estimated kinetic energy available in the collective winds of Cyg OB2 (Lozinskaya et al., 2002). The details of the modeling may be found in Butt et al. (2003).

Importantly, we find that the broadband (especially radio) emission from the secondary electrons cannot be ignored, as has often been implicitly assumed in multiwavelength analyses of hadronic gamma-ray production in SNRs, and other proposed GCR sources. This is because the age of the source (2-4×10$^6$ years) is much longer than the typical age of SNRs in their GCR acceleration phase (~10$^4$ years), and thus significantly more secondaries can accumulate in the source region (since their cooling time is longer than the few Myrs age of the source).

The spectra resulting from our calculations are presented below in figures 5, 6 for two different cases with parameters as summarized below:

- **Case I**: (predominantly hadronic generation of TeV gamma-rays) – Figure 5

    B=5$\mu$G; $E_{p\_max}$=1 PeV; $E_{e\_max}$=1 PeV; $R_{e/p}$=0.01;

    efficiency, $\eta$~ $E_{CR}/E_{kin}$ ~0.08%; density=30 cm$^{-3}$

- **Case II**: ($e^-$ IC generation of TeV gamma-rays) – Figure 6

    B=5$\mu$G; $E_{p\_max}$=1 PeV; $E_{e\_max}$=1 PeV; no protons;

efficiency, $\eta \sim E_{CR}/E_{kin} \sim 0.2\%$; density=30 cm$^{-3}$

In Figure 5 we report the scenario in which the TeV gamma-rays have a hadronic origin. The plot shows the multiband spectra from radio to gamma-ray energies due to synchrotron, bremsstrahlung and inverse Compton emission from primary electrons and secondary $e^{\pm}$, and neutral pion decay generated from p-p inelastic collisions. In this case the emission from primary electrons is shown for comparison and we assume a ratio of electrons to protons at relativistic energies of 0.01. While the TeV spectrum is well reproduced by the hadronic emission, the synchrotron emission due to secondaries generated in the same hadronic processes is below observational upper limits at both radio (1.4 GHz) and X-ray (keV range) frequencies. The predicted radio flux, in particular, is only a factor 2-3 below the observed upper-limit. With an assumed 5$\mu$G magnetic field the particles responsible for the radio-synchrotron emission at 1.4 GHz have a Lorentz factor of order 10$^4$. Given the scaling of the synchrotron emission with magnetic field as $B^{1+\alpha}$, where $\alpha$=0.5 is the spectral index, the magnetic field strength is allowed another factor two or so higher before the radio upper limit is violated.

In Figure 6 we consider the case where the TeV flux arises from electron inverse Compton emission. Thus, as compared to the previous case, we increased the injected population of electrons by a factor more than 200 (hadronic contributions are not shown here for clarity). Since the background gas density and magnetic fields are unchanged with respect to the previous case, the same description of the spectral features applies here as well. It is obvious from the figure that in this case both radio and X-ray upper limits are violated. Particularly, in order to reconcile the predicted and measured radio flux at 1.4 GHz would require a magnetic field at the level of ~1 $\mu$G, which is below the Galactic average.

Clearly, even with the low adopted magnetic field of 5$\mu$G, electrons are disfavored as the dominant source of the TeV gamma-rays since both the radio and X-ray upper-limits are violated by the synchrotron emission (Figure 5 & 6).

It is often stated that a massive and dense cloud is needed to explain the TeV emission as being hadronic in origin. However, there are two main ingredients that determine the hadronic luminosity of a given source: one is indeed the value of the ambient density, but the other is the source's local CR power. We find that the low intensity of this TeV source is easily accommodated by the combination of the empirically determined density of just ~30 nucleons cm$^{-3}$ at the source site and the ~0.1% CR acceleration efficiency (ie. ~10$^{36}$ erg s$^{-1}$ in CRs locally). There is no need to invoke a very massive and/or dense molecular cloud at the TeV source site in order to explain the multiwavelength emissions in terms of p-p interactions.

IV. Summary and conclusions

We have carried out follow-up X-ray and radio observations of the extended and steady unidentified TeV source region recently reported by the HEGRA collaboration in Cyg OB2, the most massive OB association known in the Galaxy. The new data taken together with the reexamination of archival radio, X-ray, CO, HI and IRAS data suggest that collective turbulence and large-scale shocks due to the interacting supersonic winds of the ~2600 core OB stars of Cyg OB2, with those of an outlying subgroup of powerful OB stars in Cyg OB2 are likely responsible for the observed very-high-energy gamma-ray emissions (Fig. 1). Since new analysis of 2002 HEGRA data confirm the extended and steady nature of the TeV source (Horns & Rowell, 2003), a blazar-like hypothesis of the origin of the TeV flux, such as that explored by Mukerjee et al. (2003), is now untenable. It is, however, possible that the extended TeV source is actually composed of multiple, nearby steady point-like TeV sources such as may result from a concentration of 'target' stars immersed in an intense CR bath (eg. Romero & Torres, 2003). Higher spatial resolution TeV observations, such as those made possible by HESS, may help in resolving this issue. The suggestion that the TeV source may possibly be associated with Cyg X-3 (eg. Aharonian et al., 2002) is also difficult to reconcile with the fact that Cyg X-3's jets lie at ≲14° to the line-of-sight: the de-projected distance between Cyg X-3 and the TeV source (given Cyg X-3's distance of ~9kpc) appears to be too large to support such an hypothesis.

We have carried out detailed simulations of the multifrequency spectra of the extended TeV source and favor a scenario where the TeV gamma-rays are dominantly of a nucleonic, rather than an electronic, origin. A magnetic field of just 5 $\mu$G at the TeV source site would rule against the possibility of an electronic origin of the TeV flux (Fig. 11). Since much higher fields are known to exist in young stellar associations (eg. Crutcher & Lai, 2002), a predominantly hadronic source is favored (Fig. 10). We find no need to invoke a dense and/or massive molecular cloud at the extended TeV source site to explain the multifrequency emissions in terms of accelerated hadrons.

Deeper radio and X-ray observations would be useful in order to separate the non-thermal *vs.* thermal components of the diffuse emissions so that straightforward comparisons to multiwavelength simulations can be made. A determination of the Cyg OB2 magnetic field in this region would also place strong constraints on TeV source models and is highly desirable. Further high-sensitivity infrared observations, such as those already carried out by Comerón et al. (2002), would be very useful in order to make an accurate census of the OB stars towards the highly extincted region of the extended TeV source. Of course future observations by GLAST and the next-generation of steroscopic Cherenkov telescopes (HESS, VERITAS, etc.) will be crucial in exposing the nature of this mysterious very high-energy gamma-ray source.

We are indebted to Harvey Tannenbaum and the CHANDRA X-ray Center in granting us the DDT time; and to the VLA for allowing the scheduling of the B-configuration observation on such short notice. Discussions with, and information from, Peter Biermann, Dieter Horns, Jurgen Knoldseder, Henric Krawczynski, Maura McLaughlin, Thierry Montmerle, Etienne Parizot, Jerome Rodriguez, Gavin Rowell, Diego Torres, Bulent Uyaniker, Mike Shara, David Thompson, Heinz Wendker and Dave Zurek are appreciated. YMB acknowledges the support of the *CHANDRA* project, NASA Contract NAS8-39073. Study of this extended and non-variable TeV $\gamma$-ray source is also supported by CHANDRA grant DD2-3020X. The use of the HEASARC archive at GSFC, the NASA ADS, and the Canadian Galactic Plane Survey was invaluable to this study.

**Table 1.** OB stars surrounding TeV source for $d \leq 9'^{(*)}$

| Name | R.A.$_{2000}$ | Dec$_{2000}$ | Sp.Class.$^1$ | $r$(pc) | $\log(\dot{M}_{\rm exp})^2$ |
|---|---|---|---|---|---|
| Cyg OB2-560 | 20 31 49.74 | 41 28 26.9 | O9.5 V | 2.0 | -6.564 |
| VI CYG 4 | 20 32 13.82 | 41 27 12.0 | O7 III((f)) | 1.7 | -5.567 |
| Cyg OB2-14 | 20 32 16.5 | 41 25 36 | O... | 2.4 | -6.367 |
| Cyg OB2-31 | 20 32 16.62 | 41 25 36.4 | O9 V | 2.4 | -6.367 |
| Cyg OB2-516 | 20 32 25.59 | 41 24 51.9 | O5.5 V | 3.2 | -5.432 |
| Cyg OB2-15 | 20 32 27.5 | 41 26 15 | O8 V | 2.7 | -6.099 |
| Cyg OB2-30 | 20 32 27.66 | 41 26 22.11 | O8 V | 2.7 | -6.099 |
| A43$^3$ | 20 32 38.5 | 41 25 13.0 | O... | 3.9 | -6.367 |
| MT-299$^4$ | 20 32 38.66 | 41 25 13.7 | O7.5 V | 3.9 | -6.003 |
| VI CYG 6 | 20 32 45.44 | 41 25 37.51 | O8 V: | 4.2 | -6.099 |
| NSV 13126 | 20 31 22.02 | 41 31 28.4 | B1 Ib: | 4.4 | -5.9 |
| Cyg OB2-205 | 20 31 55.9 | 41 33 04 | B1.5 V | 1.9 | -6.7 |
| Cyg OB2-210 | 20 31 56.4 | 41 31 48 | B1.5 V | 1.2 | -6.7 |
| Cyg OB2-545 | 20 32 03.3 | 41 25 12 | B0.5 :V | 2.4 | -6.4 |
| MT-213 | 20 32 12.8 | 41 22 26 | B0 Vp | 1.5 | -6.2 |
| MT-215 | 20 32 13.2 | 41 27 32 | B1 V | 1.3 | -6.6 |
| Cyg OB2-500 | 20 32 25.8 | 41 29 39 | B1 V | 1.7 | -6.6 |
| Cyg OB2-21 | 20 32 27.4 | 41 28 52 | B1 III | 1.9 | -6.0 |
| Cyg OB2-502 | 20 32 27.85 | 41 28 52 | B0.5 V | 1.9 | -6.4 |
| Cyg OB2-492 | 20 32 36.8 | 41 23 26 | B1 V | 4.3 | -6.6 |

(*): Stars selected from Simbad database, Chen et al. 1996, Comeron et al. 2002, and Massey and Tompson 1991. 1: From Simbad; 2: computed from Vink et al. 2001, if stellar luminosities, masses and effective temperatures are from Vacca et al. 1996, and terminal velocities are from Prinja et al. 1990; 3: Comerón et al. 2002; 4: Massey & Thompson 1991

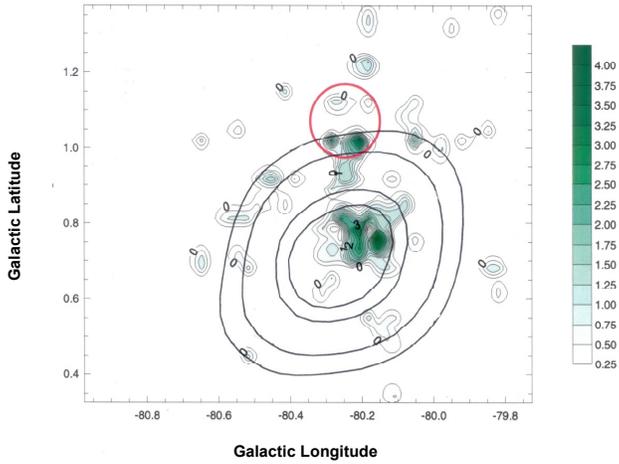

Fig. 1: Distribution of all 110 *cataloged* OB stars in Cyg OB2 shown as a surface density plot (stars per 4 arcmin$^2$). Note that many stars in Cyg OB2 remain uncataloged – the total number of OB stars alone is expected to be ~2600 (Knodlseder 2002). Although the extinction pattern towards Cyg OB2 may control the observed surface density of OB stars, our analysis assumes that the observed distribution of OB stars tracks the actual distribution. The thick contours show the location probability contours (successively, 50%, 68%, 95%, and 99%) of the non-variable MeV-GeV range EGRET $\gamma$-ray source 3EG 2033+4118 (Hartman et al., 1999). The red circle outlines the 5.6' radius extent of the diffuse and steady TeV source, TeV J2032+4131, reported by HEGRA (Rowell et al. 2002; Aharonian et al., 2002)

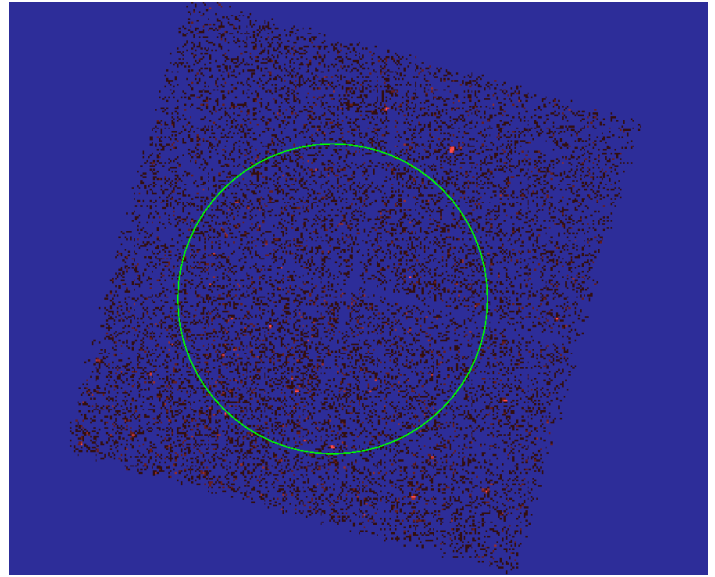

Fig. 2: The raw 5 ksec CHANDRA image of the 4 I-array chips (binned-by-8-pixels). The green circle shows the 5.6' radius extent of the diffuse TeV source, TeV J2032+4131, reported by HEGRA (Aharonian et al., 2002). The aimpoint is at the center of the circle, $\alpha_{2000}$: $20^h32^m07^s$, $\delta_{2000}$: +41°30′30″. North is up and East is to the left.

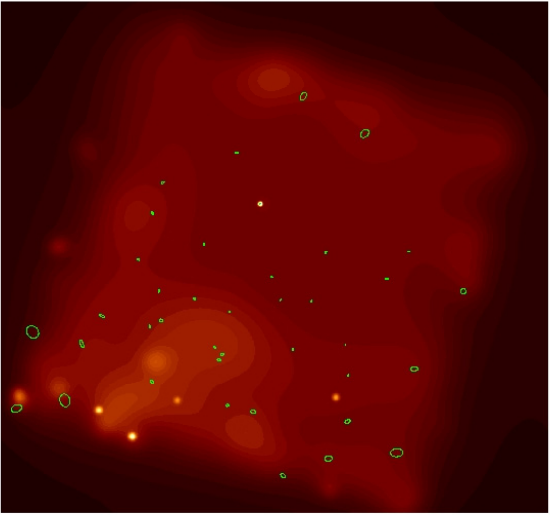

Fig. 3: An adaptively smoothed X-ray image of the TeV source region, covering the same field as in Fig. 2. The point-like sources have been removed prior to the smoothing – they are overlaid as the faint green contours. Some spurious maxima in the diffuse emission are artifacts of the smoothing algorithim. The spurious maxima are those which appear point-like, but have no true point-like (green contours) counterparts. eg. the two point-like maxima in the SE. North is up and East is to the left.

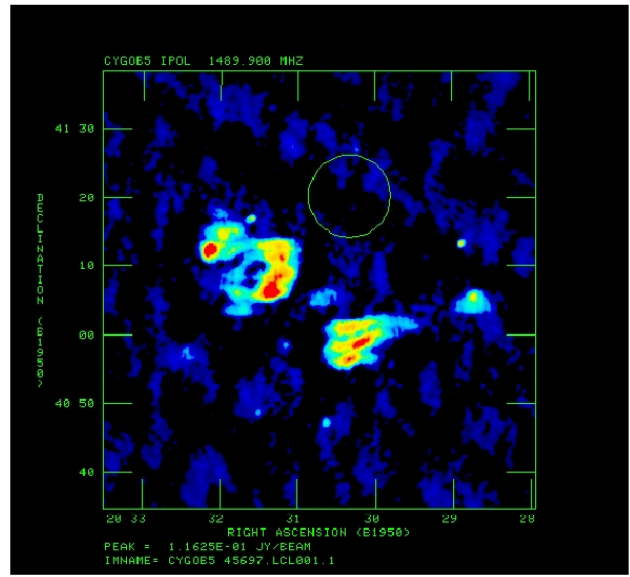

Fig. 4: The VLA D-configuration radio image of the Cyg OB2 region. The green circle shows the 5.6' radius extent of the diffuse TeV source TeV J2032+4131 reported by HEGRA (Rowell et al. 2002; Aharonian et al., 2002). The upper limit to the radio emission there at 1.49 GHz is <200 mJy.

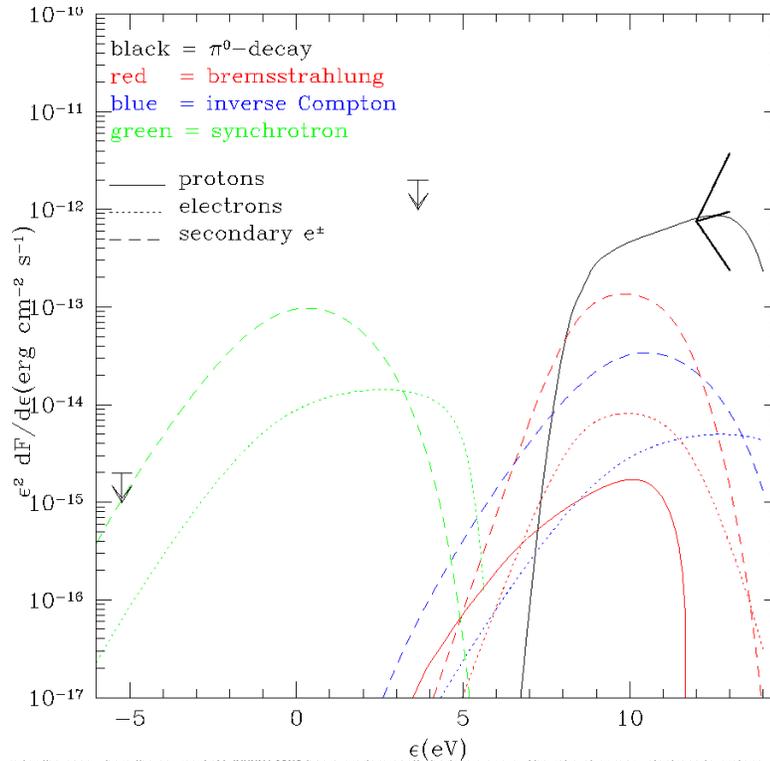

Fig. 5: A simulated multiwavelength spectrum for the case where the source TeV J2032+4131 has a predominantly hadronic origin. The ratio of primary electrons to protons was taken as 1%. A weak magnetic field of $5\mu$G was assumed, in line with the nominal Galactic value. Interestingly, the radio emission of the secondary electrons dominates the contribution from the primaries – this is because the age of the source (~2.5Myrs) exceeds the cooling time of the secondary $e^{\pm}$ and thus they simply accumulate in the source region. The injection efficiency (ratio of GCR energy to time-integrated wind power) is 0.08%. *Note that the measured X-ray flux is taken here as an upper limit to the non-thermal component alone. Deeper X-ray and radio observations will help resolve the diffuse non-thermal components, which could then be directly compared with the simulated spectrum shown here.*

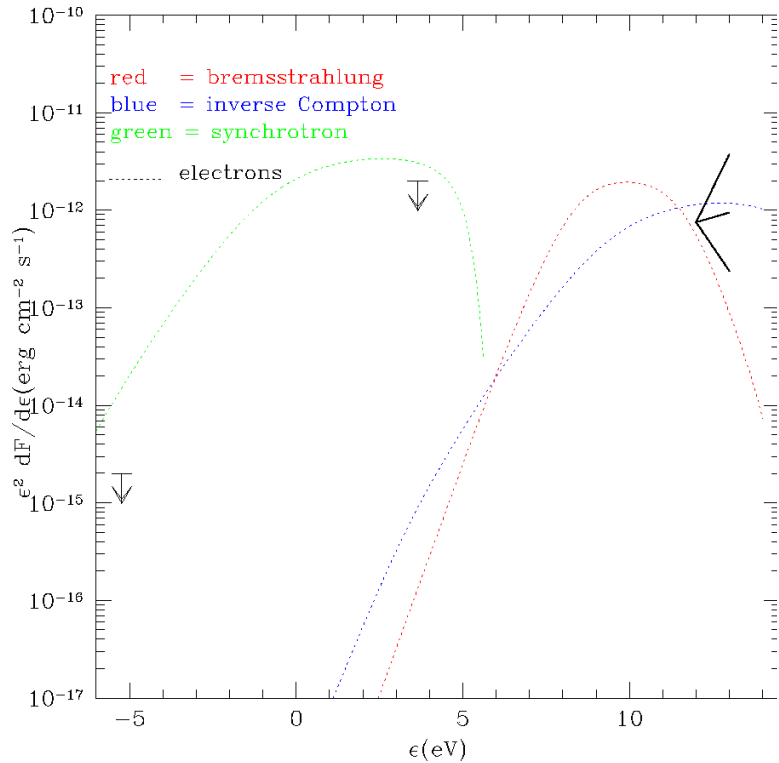

Fig. 6: A simulated multiwavelength spectrum for the case where the source TeV J2032+4131 has a purely electronic origin. A weak magnetic field of $5\mu$G was assumed, in line with the nominal Galactic value. The injection efficiency (ratio of required GCR energy to time-integrated wind power) in this case is 0.2%. *Note that since both the X-ray and radio upper limits are violated and thus an electronic origin of TeV J2032+4131 is disfavored.* If a lower magnetic field exists in the TeV source region this would, of course, decrease the synchrotron emission (green curve), and could allow for an electronic model. However, Crutcher & Lai (2002) find that magnetic fields in young star forming regions are typically even higher – and not lower – than the nominal Galactic value of $5\mu$G we have used here. Note that, as in the previous figure, the measured X-ray flux is taken here as an upper limit to the non-thermal component alone. Deeper X-ray and radio observations will help resolve the diffuse non-thermal components, which could then be directly compared with the simulated spectrum shown here.